\documentstyle[12pt]{article}

 \csname
@addtoreset\endcsname{equation}{section}

\def\a{\alpha}
\def\b{\beta}

\def\C{\Gamma}
\def\d{\delta}

\def\e{\epsilon}

\def\f{\phi}
\def\F{\Phi}

\def\l{\lambda}

\def\m{\mu}

\def\t{\tau}

\def\x{\xi}

\def\O{\Omega}

\def\cA{{\cal A}}

\def\cN{{\cal N}}

\def\cV{{\cal V}}

\def\yb{{\bar y}}
\def\zb{{\bar z}}
\def\pb{{\bar \pi}}
\def\Ah{{\widehat A}}
\def\Fh{{\widehat \F}}
\def\ad{\dot{\a}}
\def\bd{\dot{\b}}

\def\yd{{\bar y}^{\ad}}
\def\ydd{{\bar y}_{\ad}}
\def\zd{{\bar z}^{\ad}}

\def\nn{\nonumber}
\def\fr{\frac}
\def\ra{\rightarrow}
\def\lev{{(\ell)}}

\def\be{\begin{eqnarray}}
\def\ee{\end{eqnarray}}
\def\ba{\begin{array}}
\def\ea{\end{array}}
\def\bec{\begin{center}}
\def\ec{\end{center}}
\def\ns{\normalsize}
\def\ft#1#2{{\textstyle{{\scriptstyle #1} \over {\scriptstyle #2}}}}

\def\la#1{\label{#1}}
\def\eq#1{(\ref{#1})}
\let\bm=\bibitem

\newcommand{\hoch}[1]{$^{#1}$}

\begin{document}

\hfill{{CTP-TAMU-15/02}}
\\[-20pt]

\hfill{{UUITP-09/02}}
\\[-20pt]

\hfill{hep-th/0207101}

\vspace{20pt}

\begin{center}


{\Large\bf On $\cN=1,2,4$ Higher Spin Gauge Theories\\[5pt] in
Four Dimensions}


\vspace{30pt}

{\large J. Engquist\hoch1, E. Sezgin\hoch2 and
P. Sundell\hoch1}\\[20pt]

\end{center}

{\small \it \hoch1 Department of Theoretical Physics, Uppsala
University, Sweden}\\
{\small\it\hoch2 Center for Theoretical Physics, Texas A\&M University,
College Station, TX 77843, USA}

\vspace{100pt}

\begin{center}

{\large\bf Abstract}

\end{center}

We study $\cN=1,2,4$ higher spin superalgebras in four dimensions
and higher spin gauge theories based on them. We extend the
existing minimal $\cN=2,4$ theories and find a minimal $\cN=1$
theory. Utilizing the basic structure of the minimal $\cN=8$
theory, we express the full field equations for the $\cN=1,2,4$
theories in a universal form without introducing Kleinian
operators. We also use a non-minimal $\cN=4$ higher spin algebra
tensored with $U(3)$ to describe a higher spin extension of
$\cN=4$ supergravity coupled to the massless vector multiplets
arising in the KK spectrum of 11D supergravity on the $\cN=3$
supersymmetric $AdS_4\times N^{010}$ background. The higher spin
theory also contains a triplet of vector multiplets which may play
a role in the super-Higgs effect in which $\cN=4$ is broken down
to $\cN=3$.

\newpage


\section{Introduction}


Massless higher spin gauge theories are generalizations of
ordinary gravity that have anti-de Sitter spacetime as vacuum and
contain an infinite set of massless particles of ever-increasing
spin. These theories are best understood in 4D, where their
``classical'' interactions have been constructed by gauging of
higher spin algebras \cite{kv1,4dv0,4dv1,vr2}. These symmetry algebras, which
are infinite dimensional extensions of the AdS group, admit AdS
singletons as fundamental representations from which massless as
well as massive representations can be obtained by taking tensor
products. The 4D higher spin algebras also arise as symmetries of
3d theories of free $OSp(\cN|4)$ singletons. Recent developments
\cite{su2,us1,5dv1,edseminar,holo} suggest that the 4D gauge
theory with $\cN=8$ supersymmetries is a truncation of an unbroken
phase of M theory on $AdS_4\times S^7$ that has a 3d holographic
dual given by a free $SU(N)_c$-invariant singleton theory.

Given that M theory admits $AdS_4$ compactifications with
$\cN<8$ supersymmetries, it is natural to study the higher spin gauge
theory/free singleton correspondence in these cases as well. As in the
case of supergravity, the lifting of
higher spin gauge theory with $\cN<8$ to M theory requires couplings of
the minimal higher spin gauge multiplet to extra `matter' multiplets
in the form of an infinite-dimensional representation of the
minimal higher spin algebra. Since the higher spin dynamics is a highly
constrained system it is worth while to examine the nature of such
couplings. In particular, it would be interesting to study the consequences
of massless higher spins in the context of $\cN=1$ supergravity based
phenomenology.

In this paper we describe the minimal higher spin extensions of
$OSp(\cN|4)$ for $\cN=0,1,2,4,8$. The minimal $\cN=0,4,8$ theories
have appeared previously in \cite{kv1,4dv0,4dv1,us3,us4,analysis},
and the minimal $\cN=1,2$ algebras in \cite{kv0,kv1}. Consistent
interactions with $\cN=1,2$ symmetry have been given previously,
though based on slightly modified algebras with additional
generators corresponding to extra auxiliary fields \cite{mat}. We
shall elaborate on technical aspects of this point in the
conclusion of Section 3, and the introduction and conclusion of
Section 6. In this paper we also construct non-minimal $\cN=2,4$
theories and exhibit various relationships between $\cN\leq 8$
theories by means of consistent truncations starting from the
minimal $\cN=8$ theory. As for $\cN=3$ models, we shall comment
briefly on them below. A summary of the ${\cN=0,1,2,4,8}$ higher
spin algebras discussed in this paper is given in Table
\ref{summary} where different notation used for them in the
literature is also exhibited.

The higher spin algebras considered in this paper contain
supercharges which are realized using fermionic Clifford
elements transforming as vectors of the R-symmetry group
$SO(\cN)_R$, and in some cases, most notably $\cN=1$ and $\cN=2$,
an additional set of fermionic Clifford elements. These
elements are the real and imaginary parts of the fermionic
creation and annihilation operators used to build the
underlying $OSp(\cN|4)$ singleton Fock spaces, as summarized
in Table \ref{summary}. Besides being a natural construction
of the higher spin superalgebras, this formulation enables a
universal treatment of the full field equations as well as the
algebraic conditions on the higher spin master fields. The
distinction between the various theories lies in a certain
chirality operator $\C$, which is the product of all fermionic
Clifford elements.

It is worth pointing out that none of the oscillator realizations
of the higher spin algebras discussed in this paper employ
Kleinian operators \cite{valg,vpro}, which are square-roots
of the unit operator that anti-commute with the basic
bosonic oscillator.
In the case of $\cN=1,2$ the addition of Kleinian
operators imply enlarged algebras that contain extra generators
\cite{valg,vpro} that correspond to extra auxiliary fields upon gauging
\cite{mat}. These do not affect the spectrum, but mediate
additional interactions. The formulations without Kleinian
operators considered here in the case of $\cN=1,2$ are therefore
more restrictive, and as a result they admit less general
interaction ambiguities, as we shall discuss in more detail in
Section 6.

The consistency of higher spin gauge theories implies that the
spectrum of massless physical fields form unitary representations
of the higher spin algebras. We are interested in higher spin
gauge theories whose massless spectra arise in the product of two
singletons which in turn form a fundamental UIR of the higher spin
algebra. This fits naturally into the framework of holography,
whereby the massless bulk fields couple to bilinear operators
formed out of free singleton fields in 3d. In this paper we
determine the exact form of the singleton products in all cases,
as summarized in Table \ref{summary}.

The $\cN=3$ case is not listed because the $OSp(3|4)$ singleton is
also a UIR of $OSp(4|4)$. Thus the spectra of massless fields
obtained from tensoring two singletons in two cases are the same.
However, coupling the $\cN=4$ theory to a suitable Higgs sector,
one may find a vacuum in which the higher spin symmetries break
down to the symmetries of a matter coupled $\cN=3$ supergravity.

In the case of matter coupling to higher spin gauge theories for
$\cN\leq 4$, it is natural to introduce flavor symmetry through
Vasiliev's matrix valued higher spin algebras \cite{mat,kv1}. The
precise form of the matter coupling required for uplifting to M
theory on $AdS_4 \times X^7$ requires a knowledge of the
kinematics of the corresponding 3d SCFT, which is determined to a
large extent from the geometry of the conifold over $X^7$
\cite{klebanov,n2,v52,n3}. In this paper we discuss the matter
couplings in the $\cN=4$ higher spin theory relevant for M theory
on $AdS_4\times N^{010}$, which is the only known smooth $X^7$
with $\cN=3$ \cite{warner}. Interestingly enough, we find a
triplet of $\cN=4$ vector multiplets which may play a role in the
super-Higgs effect in which $\cN=4$ is broken down to $\cN=3$
\cite{shadow}. We also identify the Betti vector multiplet and the
$SU(3)$ Yang-Mills multiplet which arise in the massless KK
spectrum of the 11D supergravity on $AdS_4\times N^{010}$
\cite{n010}.

The paper is organized as follows: In Section 2 we review the
minimal bosonic theory. The $\cN=1,2,4$ higher spin algebras and
their massless spectra are described in Sections 3,4,5,
respectively. Full field equations and consistent truncations are
discussed in Sections 6 and 7, respectively. In Section 8 we
discuss a matter coupled $\cN=4$ higher spin gauge which may
describe the massless sector of an unbroken phase of M theory on
$AdS_4\times N^{010}$.


\section{The Minimal Bosonic Theory}


In order to explain the main ideas \cite{4dv1} and to set our
notation \cite{analysis} we first review the minimal bosonic
theory based on the higher spin algebra\footnote{This algebra
is isomorphic to the algebra $hs_2(1)$ discussed in \cite{kv0}.}

\be hs(4)\supset SO(3,2)\ .\ee

Physical UIRs of $SO(3,2)$ are lowest weight representations
denoted by $D(E_0,s)$ where
$E_0$ and $s$ are the energy and the spin, respectively,
carried by the lowest weight states. The
energy and the spin are defined by the generators of the
maximal compact subalgebra
$SO(3)\times SO(2)_E\subset SO(3,2)$. Unitarity requires either
$E_0\geq s+1$, $s=0,1/2,1,3/2,\ldots$ or $E_0=s+1/2$, $s=0,1/2$.
The representations saturating the bounds of the continuous
series, i.e. $E_0=s+1$, and the special case $(E_0,s)=(2,0)$
are the massless representations. For $s\geq 1$ these describe
tensors in AdS$_4$ obeying gauge invariant field equations.
The special UIRs $D(\ft12,0)$ and $D(1,\ft12)$ are known as singletons
and describe conformal scalars and fermions, respectively, in $d=3$.
Importantly, as we shall see below, the singletons are
UIRs also of the higher spin algebra $hs(4)$.

The spectrum of massless physical fields of
the $hs(4)$ higher spin gauge theory is given by

\be \nn {\rm Spectrum}\left[hs(4)\right]&=&
\left(D(\ft12,0)\otimes
D(\ft12,0)\right)_{\rm S}\\[10pt]&=&D(1,0)+ D(3,2)+ D(5,4) + D(7,6)+\cdots\ .
\la{eq:uirb} \ee

Thus the minimal spectrum
consists of real massless fields with
spin $s=0,2,4,\ldots$, each occurring once. The minimal theory is the
consistent truncation of the non-minimal bosonic theory described
in \cite{vr1}\footnote{The
non-minimal theory described in \cite{vr1} is obtained by relaxing
the $\t$ conditions given in \eq{eq:conboson1} and
\eq{eq:conboson2} to $\t^2(A)=A$ and $\t^2(\F)= \F$. An
alternative formulation based on linear $\t$ conditions is given
at the end of Section 7; see \eq{alt1}, \eq{alt2}, \eq{alt3} and
\eq{alt4}. The resulting higher spin algebra has maximal
finite-dimensional subalgebra $SO(3,2)\times U(1)_f$, where $f$
denotes flavor, and the spectrum of the theory is neutral under
$U(1)_f$. We therefore denote the higher spin algebra by
$hs_0(4;1)$, where $(4;1)$ refers to $Sp(4)\times U(1)_f$ and the
subscript denotes that all $U(1)_f$ charges vanish.} whose spectrum include also the real massless fields with
spin $s=1,3,5,\ldots$ that are contained in the anti-symmetric product

\be \left(D(\ft12,0)\otimes D(\ft12,0)\right)_{\rm A}=D(2,1)+
D(4,3)+ D(6,5) + D(8,7)+\cdots\ . \la{anti} \ee

The fact that the tensor product of $3d$ singletons gives rise to a massless
spectrum in $AdS_4$ was first observed by Flato and Fronsdal \cite{fronsdal}.
The
occurence of infinitely many lowest weight states in the
tensor product stems from the fact that the singletons are infinite
dimensional representations and that the space of states in the
tensor product with fixed energy, which is a finite-dimensional space,
containts precisely one massless lowest weight state. To
give these states one can make use of the Jordan
decomposition $SO(3,2)=
L^-\oplus L^0\oplus L^+$, where $L^0=SO(3)\times SO(2)_E$ is spanned
by generators $L^I_J$ where $I,J=1,2$, and $L^+$ and $L^-$ are
spanned by generators $L^{IJ}$
and $L_{IJ}$, respectively. The scalar singleton weight space is given by
$D(\ft12,0)=\{L^{(I_1J_1}\cdots L^{I_nJ_n)}|\ft12,0\rangle\}_{n=0}^\infty$,
and the massless lowest weight state with energy $E_0=s+1$ in
$D(\ft12,0)\otimes D(\ft12,0)$ is given by

\be |s+1,s\rangle =\sum_{k=0}^s  (-1)^k {s\choose k}
L^{(I_1J_1}_{(1)}\cdots L^{I_kJ_k}_{(1)}L^{I_{k+1}J_{k+1}}_{(2)}
\cdots L^{I_sJ_s)}_{(2)}|\ft12,0\rangle_{(1)} \otimes
|\ft12,0\rangle_{(2)}\ ,\phantom{aaa}\ee

where $|\ft12,0\rangle_{(\a)}$ ($\a=1,2$) denote the
ground states of the two copies
of the singleton making up the tensor product, and $L^{IJ}_{(\a)}$
denote the energy-raising operators acting in these two spaces.

The minimal bosonic algebra $hs(4)$ can be realized using a
set of Grassmann even spinor oscillators $y_{\a}$ and
$\ydd=(y_{\a})^{\dagger}$ obeying the oscillator algebra

\be
y_\a\star y_{\b}=y_\a y_\b+i\e_{\a\b}\ ,\qquad
\ydd\star\yb_{\bd}=\ydd\yb_{\bd}+i\e_{\ad\bd}\ ,
\ee

where the $y_\a\star y_\b$ denotes the operator product and $y_\a
y_\b= y_\b y_\a$ denotes the Weyl ordered product. The operator product
product gives rise to an associative algebra $(\cA,\star)$ of arbitrary
Weyl ordered polynomials in the oscillators. The algebra $(\cA,\star)$
has an anti-involution
$\t$ and involutions $\pi$ and $\bar \pi$ with the following
action on a Weyl ordered function $f\in \cA$:

\be
\tau(f(y,\yb))=f(iy,i\yb)\ ,\qquad
\pi(f(y,\yb))=f(-y,\yb)\ ,\qquad
\pb(f(y,\yb))=f(y,-\yb)\ .\la{taupi} \ee

Note that $\tau(f\star g)=\tau(g)\star\tau(f)$ and $\pi(f\star
g)=\pi(f)\star\pi(g)$. The $hs(4)$ gauge theory is constructed by
introducing a one-form gauge field $A=dx^\m A_{\m}(x;y,\yb)$ and a
zero-form $\F(x;y,\yb)$ satisfying the conditions:

\be
\la{eq:conboson1} \tau(A)&=&-A\ ,\qquad~~ A^{\dagger}~=~-A\ ,\\[10pt]
\la{eq:conboson2} \tau(\F)&=&\pb(\F)\ ,\qquad
\F^{\dagger}~=~\pi(\F)\ .\ee

The conditions on $A$ and $\F$ define the adjoint and
quasi-adjoint (also known as the twisted adjoint \cite{vr2})
representations of $hs(4)$, respectively. Thus
$hs(4)$ consists of arbitrary Weyl ordered polynomials $P(y,\yb)$
obeying the conditions in \eq{eq:conboson1}. The Lie bracket
is given by $[P_1,P_2]=P_1\star P_2-P_2\star P_1$, which
closes by virtue of \eq{taupi} and the fact that the
hermitian conjugation acts as an (anti-linear) anti-involution.
The quasi-adjoint representation of $hs(4)$ is
given by $P(\Phi)=P\star \Phi-\Phi\star \bar\pi(P)$.

The conditions on $\F$ and $A$ are engineered such that the
component fields obtained by expansion in $y$ and $\yb$ are of
the correct form for writing linearized field equations giving
rise to the spectrum \eq{eq:uirb} upon gauging. To see this we
begin by writing the general solution to \eq{eq:conboson1} and
\eq{eq:conboson2} as follows:

\be A_\m=\fr1{2i}\sum_{\ell=0}^\infty A_\m^{(\ell)}\ ,\qquad
\F=\sum_{\ell=-1}^\infty \F^{(\ell)}\ ,\ee

where the $\ell$th level is given by

\be A^{(\ell)}_\m =\sum_{m+n=2\ell+2} A^{(\ell)}_\m(m,n)\ ,\qquad
\F^{(\ell)}=\sum_{|m-n|=2\ell+2}\F^{(\ell)}(m,n)\ .\ee

We use the short-hand notation

\be f(m,n) = \fr1{m!n!}y^{\a_1}\cdots y^{\a_m}\yb^{\ad_1} \cdots
\yb^{\ad_n}f_{\a_1\ldots\a_m\ad_1\ldots\ad_n}(x)\ .\la{not}\ee

The gauge field $A^{(0)}_\m$ contains the vierbein and the Lorentz
connection, which gauge the $SO(3,2)$ generators
$P^{\a\ad}=y^{\a}\yd$ and $M^{\a\b}=y^{\a}y^{\b}$. Thus
$hs(4)\supset SO(3,2)$, which is the maximal finite-dimensional
subalgebra. By splitting $y_\a$ into creation operators $a_I$
 ($I=1,2)$ and annihilation operators
$a^I=(a_I)^\dagger$,
for example by writing

\be y_1=a_1+i(a_2)^\dagger\ ,\qquad y_2=-a_2+i(a_1)^\dagger\ ,\la{bososc}\ee

the algebra $\cA$
can be represented unitarily in a Fock space. The space of
states built from an even number of oscillators is isomorphic to $D(\ft12,0)$
and that built from an odd number of oscillators is isomorphic to
$D(1,\ft12)$. Since $hs(4)$ consists of even polynomials in oscillators
it follows that the $SO(3,2)$ singletons, and hence the tensor
products in \eq{eq:uirb} and \eq{anti}, are UIRs of
$hs(4)$.

The full field equations for the non-minimal bosonic theory based
on $hs_0(4;1)$ were given in \cite{vr1} and their minimal truncation
based on $hs(4)$ was analyzed in more detail in \cite{analysis}.
The linearized field equations are obtained
by expanding around an AdS background by setting $A=\O+W$, where
$d\O+\O\star \O=0$, and treating $W$ and $\F$ as small fluctuations.
The resulting linearized equations are

\be \la{lc1}dW^\lev +\O\star
W^\lev+W^\lev\star\O&=&e^{\a\ad}\wedge
e^{\b}{}_{\ad}{\partial^2\Phi^\lev
(y,0)\over\partial y^{\a}
\partial y^{\b}}-{\rm h.c.}\ ,\quad \ell\geq 0\ ,\\[10pt]
\la{lc2}d\F^\lev +\O\star \F^\lev-\F^\lev\star\pb(\O)&=&0\ ,
\qquad\qquad\qquad\qquad\qquad \ell\geq -1\ .\ee

It follows from these equations that all component fields can be
expressed in terms of a physical scalar residing at level
$\ell=-1$ in $\F$ and physical spin $s=2,4,\ldots$ fields
at levels $\ell=(s-2)/2$ in $A_\m$. From
\eq{lc1} it follows that $\F^\lev(2\ell+2,0)$, $\ell\geq 0$ is
the spin $s=2\ell+2$ Weyl tensor, which is given as $s$
derivatives of the physical spin $s$ gauge field. Moreover, from \eq{lc2}
it follows that $\F^\lev(2\ell+2+k,k)$, $\ell\geq -1$, $k>0$ are
$k$ derivatives of the leading component $\F^\lev(2\ell+2,0)$.
Thus the complete spectrum of massless physical fields can be deduced from the
expansion of the master scalar field $\Phi$ alone.


\section{The Minimal $\cN=1$ Theory}


In this section we construct a minimal $\cN=1$ higher spin gauge
theory based on the higher spin algebra\footnote{This algebra
is isomorphic to $shs^f(1|0)$ given in \cite{kv0}.} $hs(1|4)$ whose maximal
finite-dimensional subalgebra is given by

\be OSp(1|4)\subset hs(1|4)\ .\ee

The spectrum of the $ hs(1|4)$ theory can be obtained by considering
the tensor product of two $OSp(1|4)$ singletons

\be \Xi=D(\ft12,0)+D(1,\ft12)\ .\la{singn1}\ee

The tensor products of $D(\ft12,0)$ and $D(1,\ft12)$ are given by

\be \nn D(\ft12,0)\otimes D(\ft12,0)&=&\left[D(1,0)+ D(3,2)+
D(5,4)+ \cdots \right]_{\rm S}\\
&&+ \left[D(2,1)+ D(4,3)+
D(6,5)+ \cdots \right]_{\rm A}\\[10pt]
\nn D(1,\ft12)\otimes D(1,\ft12)&=&\left[D(2,1)+ D(4,3)+
D(6,5)+ \cdots \right]_{\rm S}\\
&&+\left[D(2,0)+ D(3,2)+
D(5,4)+ \cdots \right]_{\rm A}\\[10pt]
D(\ft12,0)\otimes D(1,\ft12)&=&D(\ft32,\ft12)+ D(\ft52,\ft32)+
D(\ft72,\ft52)+ \cdots \ .\ee

The spectrum of the $hs(1|4)$ theory is defined by the symmetric tensor product

\be \la{specn1} {\rm Spectrum}\left[hs(1|4)\right]=(\Xi\otimes
\Xi)_{\rm S}\ ,\ee

which is tabulated in Table \ref{spec1}. In addition to the supergravity
multiplet, the spin $s\leq 2$
sector of the spectrum contains a scalar multiplet at level $(-1,1/2)$.
In the full spectrum each even spin occurs twice, odd spins do not
occur and each half-integer spin occurs once.

The minimal $\cN=1$ theory is obtained by starting from the
bosonic theory described in the previous section and adding
two fermionic oscillators $\xi$ and $\eta$ obeying:

\be \xi\star \xi=1\ ,\qquad \xi\star\eta=\xi\eta=-\eta\xi\ ,\qquad \eta\star
\eta=1
\ .\ee

Involutions $\pi$ and $\bar \pi$ and a graded anti-involution $\t$ are defined
by

\be \t(f(y,\yb,\xi,\eta))&=&f(iy,i\yb,i\xi,-i\eta)\ ,\la{tau1}\\
\pi(f(y,\yb,\xi,\eta))&=&f(-y,\yb,\xi,\eta)\ ,\la{pi1}\\
\bar \pi(f(y,\yb,\xi,\eta))&=&f(y,-\yb,\xi,\eta)\ ,\la{pibar1}\ee

where $\t$ acts as

\be \t(f\star g)=(-1)^{\e(f)\e(g)}\t(g)\star \t(f)\ ,\ee

and $\e(f)$ denotes the Grassmann parity of $f$ which is $0$
for bosons and $1$ for fermions.
The $hs(1|4)$-valued gauge field $A$ and the corresponding quasi-adjoint master
field $\F$ are defined by

\be
\la{an1} \tau(A)&=&-A\ ,\qquad~~ A^{\dagger}~=~-A\ ,\\
\la{fn1} \tau(\F)&=&\pb(\F)\ ,\qquad
\F^{\dagger}~=~\pi(\F)\star \C\ ,\ee

where

\be \C&\equiv& i\x\eta\ ,\\ \quad \tau(\C)&=&\C^\dagger=\C\ ,\qquad
\C\star\C=1\ ,\la{Gamma}\ee

and by Grassmann parities

\be \e(A)=\e(\F)=0\ ,\la{grassmann}\ee

which ensures the correct spin-statistics relation for the component
fields. The $\cN=1$ master fields can be expanded as:

\be A_\m&=&\sum_{\ell=0}^\infty \left(A_\m^{(\ell,0)}+
A_\m^{(\ell,1/2)}\right)\ ,\\
\qquad \F&=&\F^{(-1,1/2)}+\sum_{\ell=0}^\infty
\left(\F^{(\ell,0)}+\F^{(\ell,1/2)}\right)\ ,
\ee

where $A_\m^{(\ell,j)}$ and $\F^{(\ell,j)}$ are given by ($j=0,1/2$)

\be
A_\m^{(\ell,j)}&=&\sum_{m+n+p=4\ell+2+2j}
A^{(\ell,j)}_{\m,p}(m,n)\xi^p\eta^{2j}\ ,\\[10pt]
\F^{(\ell,j)}&=&C^{(\ell,j)}+\pi\Big(C^{(\ell,j)}{}^\dagger\Big)\star
\C
\ ,\\[10pt]
C^{(\ell,j)}&=&\sum_{n-m-p=4\ell+2j+3}
\F^{(\ell,j)}_p(m,n)\xi^p\eta^{1-2j}\ .\ee

In particular, the scalar multiplet arises at level $(-1,1/2)$ as

\be
C^{(-1,1/2)}= \phi +\yb^{\ad}\xi \l_{\ad}+\cdots\ ,\ee

where the omitted terms are derivatives of the physical fields.

The gauge field $A^{(0)}_\m$ gauges the $OSp(1|4)$ generators
$P^{\a\ad}$, $M^{\a\b}$ and $Q_\a=y_\a\x$. The unitary
Fock space realization of $hs(1|4)$ is obtained from \eq{bososc} and
the fermionic creation operator

\be \psi^\dagger=\xi+i\eta\ .\ee

The singleton $\Xi$ is isomorphic to the space of states built from a
total even number of oscillators. Since $hs(1|4)$ consists of even
polynomials in oscillators, it follows that $\Xi$ and hence the massless
spectrum \eq{specn1} are UIRs of $hs(1|4)$.

In Section \ref{Sec:full} we describe the full field equations for the $\cN=1$
theory. The linearized form of these equations is given by ($j=0,1/2$):

\be \nn dW^{(\ell,j)} +\O\star
W^{(\ell,j)}+W^{(\ell,j)}\star\O&=&
e^{\a\ad}\wedge e^{\b}{}_{\ad}{\partial^2\Phi^{(\ell,j)} (y,0,\xi,\eta)
\over\partial y^{\a}
\partial y^{\b}}\star\C\\&&\nn -e^{\ad\a}\wedge e^{\bd}{}_{\a}
{\partial^2\Phi^{(\ell,j)} (0,\yb,\xi,\eta)
\over\partial \yb^{\ad}
\partial \yb^{\bd}} \ ,\qquad \ell\geq 0\ ,\\[10pt]
\la{lcn12}d\F^{(\ell,j)} +\O\star \F^{(\ell,j)}-\F^{(\ell,j)}
\star\pb(\O)&=&0\ ,
\qquad
\ell\geq -1\ . \ee

The physical fields arising at level $(\ell,j)$ precisely correspond to
those given in Table \ref{spec1}.

We conclude by noting that an important ingredient of the $\cN=1$
theory constructed above is the $\eta$ oscillator which is required
for the existence of a $\C$ operator obeying \eq{Gamma},
which in turn is crucial for the existence
of non-trivial dynamics (see Section 6).

Instead of introducing the $\eta$ oscillator, it is also possible
to construct consistent interactions with $\cN=1$ by making use of
Kleinian operators \cite{mat,4dv0,4dv1}. This results, however, in
an enlargement of the algebra corresponding to extra auxiliary fields
in the gauge theory (which do not affect the spectrum).
It may be possible to consistently truncate
the enlarged $\cN=1$
theory down to the $hs(1|4)$ theory considered
here, as will be discussed in Section 6.

{\footnotesize
\tabcolsep=1mm
\begin{table}[t]
\bec
\begin{tabular}{|l|cccccccccccl|}\hline
& & & & & & & & & & & & \\
{${(\ell,j)}\backslash s$} & $0$ & \ns{$\ft12$} & $1$
& \ns{$\ft32$} & $2$ & \ns{$\ft52$} &
$3$ & \ns{$\ft72$} & $4$ & \ns{$\ft92$} & $5$ & $\cdots$
\\
& & & & & & & & & & & & \\
\hline
& & & & & & & & & & & & \\
$(-1,1/2)$ & $1\!+\!{\bar 1}$ & $1$ & & & & & &
& & & & \\
$(0,0)$ & & & & $1$ & $1$ & & & & &
& & \\
$(0,1/2)$ & & & & & $1$ & $1$ & & & &
& & \\
$(1,0)$ & & & & & & & & $1$ & $1$ &
& & \\
$(1,1/2)$ & & & & & & & & & $1$ & $1$
& & \\
\phantom{aa}$\vdots$ & & & & & & & & & & & &
 \\ \hline
\end{tabular}
\ec \caption{{\small  The spectrum of massless physical fields of
the minimal $\cN=1$ theory arranged into levels of $\cN=1$
multiplets labelled by $(\ell,j)$ with $s_{\rm max}=2\ell+2+j$.}}
\label{spec1}
\end{table}}


\section{$\cN=2$ Theories}


In this section we first describe a slight reformulation, without
Kleinian operators and with a truncation of certain auxiliary
fields, of the minimal $\cN=2$ theory originally constructed in
\cite{4dv1,vr2}. We then give
a non-minimal $\cN=2$ theory, again without
Kleinian operators, containing fields carrying an extra $U(1)_f$
charge. In both cases the linearized field equations take the
same form as in linearized $\cN=1$ equations
\eq{lcn12} with the proper definition of the
$\C$-operator. The full field equations and consistent truncations
are discussed in Sections \ref{Sec:full} and \ref{Sec:cons}. In both
cases the physical field contents of the master fields
are in one-to-one correspondence with
the corresponding spectra; see Tables \ref{spec2} and \ref{spec2b}.


\subsection{The Minimal $\cN=2$ Theory}


The minimal $\cN=2$ higher spin gauge theory in $4D$ is based on
the higher spin algebra which we denote by\footnote{In \cite{kv1}
this algebra is denoted by $shs^E(2,4)\simeq hu(1;1|4)$.}
$hs_0(2|4;1)$ whose
maximal finite-dimensional subalgebra is given by \cite{kv1,vr2}

\be OSp(2|4)\times U(1)_f\subset hs_0(2|4;1)\ .\ee

The R-symmetry group is $U(1)_R\subset OSp(2|4)$.
The $OSp(2|4)$ singleton is given by

\be \la{xi2} \Xi=\Xi_1+\overline{\Xi}_{-1}\ ,\ee

where

\be \Xi_1=D(\ft12,0;1,1) + D(1,\ft12;-1,1)\ee

and we use the notation $D(E,s;r,q)$ in which
$r$ and $q$ denote the $U(1)_R$ and $U(1)_f$ charges, respectively.
The spectrum of the $hs_0(2|4;1)$ theory is given by the
$U(1)_f$ neutral sector of the symmetric product of
two $OSp(2|4)$ singletons

\be \la{eqspec2} {\rm
Spectrum}\left[hs_0(2|4;1)\right]=\left(\Xi_1\otimes\overline\Xi_{-1}
\right)_{\rm S}\ ,\ee

which is tabulated in Table \ref{spec2}. In addition to
the supergravity multiplet, the spin $s\leq 2$ sector of the
spectrum contains a vector multiplet (but no hypermultiplet).

The minimal $\cN=2$ theory is obtained by starting from the
bosonic theory described in Section 2 and adding two sets of real
fermionic oscillators $\xi^r$ and $\eta^i$ $(r,i=1,2)$ obeying

\be \la{fosc2} \xi^r\star \xi^s=\xi^r\xi^s+ \d^{rs}\ ,\qquad
\xi^r\star\eta^i=\xi^r\eta^i\ ,\qquad \eta^i\star \eta^j=\eta^i \eta^j+\d^{ij}
\ .\ee

The indices $r$ and $i$ label $U(1)_R$ and $U(1)_f$ doublets, respectively.
The maps $\t$, $\pi$ and $\bar \pi$ are defined
as in \eq{tau1}, \eq{pi1} and \eq{pibar1}, and the master
fields $A$ and $\F$ as in \eq{an1}, \eq{fn1} and \eq{grassmann} where now

\be \C=\C_\x\C_\eta\ ,\qquad \C_\x=i\x^1\x^2\ ,
\quad \C_\eta=i\eta^1\eta^2\ ,\ee

together with the $U(1)_f$ invariance conditions

\be \la{u1f} [\C_\eta,A]_{\star}=0=[\C_\eta,\F]_{\star}\ .
\ee

The resulting $hs_0(2|4;1)$ master fields can be expanded as:

\be A_\m&=&A_\m^{(-1,1)}+\sum_{\ell=0}^\infty \left(A_\m^{(\ell,0)}+
A_\m^{(\ell,1)}\right)\ ,\la{a2e}\\
\qquad \F&=&\F^{(-1,1)}+\sum_{\ell=0}^\infty
\left(\F^{(\ell,0)}+\F^{(\ell,1)}\right)\ ,
\ee

where $A_\m^{(\ell,j)}$ and $\F^{(\ell,j)}$ are given by ($j=0,1$)

\be
\la{alj2} A_\m^{(\ell,j)}&=&\sum_{m+n+p=4\ell+2+2j}
A^{(\ell,j)}_{\m,r_1\dots r_p}(m,n)\xi^{r_1}\cdots \xi^{r_p}
(\C_\eta)^{j}\ ,\\[10pt]
\la{flj2a}
\F^{(\ell,j)}&=&C^{(\ell,j)}+\pi\Big(C^{(\ell,j)}{}^\dagger\Big)
\star \C\ ,\\[10pt]
\la{alj2b} C^{(\ell,j)}&=&\sum_{n-m-p=4\ell+2j+2}
\F^{(\ell,j)}_{r_1\dots r_p}(m,n)\xi^{r_1}\cdots \xi^{r_p}(\C_\eta)^{1-j}\ .\ee

In particular, the vector multiplet arises at level $(-1,1)$ as

\be
C^{(-1,1)}=\phi +\yb^{\ad}\xi^r\l_{\ad\,r}
+\yb^{\ad} \yb^{\bd} \C_\x F_{\ad\bd}+\cdots\ ,\ee

where the omitted terms are derivatives of the physical fields.

The gauge field $A^{(0)}_\m$ gauges the $OSp(2|4)$ generators
$P^{\a\ad}$, $M^{\a\b}$, $Q^r_\a= y_\a\x^r$ and $\C_\x$.
The $U(1)_f$ generator $\C_\eta$ is gauged by $A_\m^{(-1,1)}$.
The $U(1)_R$ gauge coupling is given by the ratio between the
Planck length and the AdS radius, which are the only
constants in the theory.

The unitary Fock space realization of $hs_0(2|4;1)$ is obtained
from \eq{bososc} and the fermionic creation operators

\be \psi^{1\,\dagger}=\xi^1+i\eta^1\ ,\qquad
\psi^{2\,\dagger}=\xi^2+i\eta^2 .\ee

The singleton $\Xi$ is isomorphic to the space of states built
from an even number of oscillators. Since $hs_0(2|4;1)$ consists of
even polynomials in oscillators, it follows that $\Xi$ and hence
$(\Xi\otimes \Xi)_{\rm S}$ are unitary representations of
$hs_0(2|4;1)$. Due to the $U(1)_f$ invariance condition \eq{u1f} on
the $hs_0(2|4;1)$ generators, the spectrum \eq{specn1}
given in Table \ref{spec2} forms a UIR of $hs_0(2|4;1)$.

{\footnotesize
\tabcolsep=1mm
\begin{table}[t]
\bec
\begin{tabular}{|l|cccccccccccl|}\hline
& & & & & & & & & & & & \\
{$(\ell,j)\backslash s$} & $0$ & \ns{$\ft12$} & $1$
& \ns{$\ft32$} & $2$ & \ns{$\ft52$} &
$3$ & \ns{$\ft72$} & $4$ & \ns{$\ft92$} & $5$ & $\cdots$
\\
& & & & & & & & & & & & \\
\hline
& & & & & & & & & & & & \\
$(-1,1)$ & $1\!+\!{\bar 1}$ & $2$ & $1$ & & & &
& & & & & \\
$(0,0)$ & & & $1$ & $2$ & $1$ & & & & &
& & \\
$(0,1)$ & & & & & $1$ & $2$ & $1$ & & &
& & \\
$(1,0)$ & & & & & & & $1$ & $2$ & $1$ &
& & \\
$(1,1)$ & & & & & & & & & $1$ & $2$
& $1$ &  \\
\phantom{aa}$\vdots$ & & & & & & & & & & & &
 \\ \hline
\end{tabular}
\ec \caption{{\small The spectrum of massless physical fields of
the minimal $\cN=2$ theory based on $hs_0(2|4;1)$ arranged into
levels of $\cN=2$ supermultiplets labelled by $(\ell,j)$ and with
$s_{\rm max}=2\ell+2+j$.}} \label{spec2}
\end{table}}


\subsection{A Non-minimal $\cN=2$ Theory with $U(1)_f$ Flavor}


There exists a natural generalization of the minimal $\cN=2$
theory described above in which the $U(1)_f$ invariance condition
\eq{u1f} is dropped. The resulting higher spin algebra, which we
denote by
\footnote{An extension of this algebra that includes an infinite
set of auxiliary fields has been constructed in \cite{fv1}, where
it is called $shsa(1)$, and its subalgebras and linearized field
equations are studied. A superspace action formulation of the free
theory based on $shsa(1)$ has been provided in \cite{gks1}.}
$hs(2|4;1)$, has the same maximal finite-dimensional subalgebra,

\be OSp(2|4)\times U(1)_f\subset hs(2|4;1)\ ,\ee

and
singleton representation $\Xi$ given in \eq{xi2},
as does $hs_0(2|4;1)$. The non-minimal spectrum is given by

\be {\rm Spectrum}\left[hs(2|4;1)\right]=(\Xi\otimes\Xi)_{\rm S}\
,\ee

which is listed in Table \ref{spec2b}. The
$hs(2|4;1)$ master fields have the expansions:

\be A_\m&=&A_\m^{(-1,1)}+\sum_{\ell=0}^\infty \left(A_\m^{(\ell,0)}+
A_\m^{(\ell,1/2)}+A_\m^{(\ell,1)}\right)\ ,\\
\qquad \F&=&\F^{(-1,1/2)}+\F^{(-1,1)}+
\sum_{\ell=0}^\infty \left(\F^{(\ell,0)}+\F^{(\ell,1/2)}+
\F^{(\ell,1)}\right)\ ,
\ee

where $A_\m^{(\ell,j)}$ and $\F^{(\ell,j)}$ are given by \eq{alj2},
\eq{flj2a} and \eq{alj2b} for $j=0,1$ and by the following expansion
for $j=1/2$

\be
A_\m^{(\ell,1/2)}&=&\sum_{m+n+p=4\ell+3}
A^{(\ell,1/2)}_{\m\,r_1\dots r_p\,i}(m,n)\xi^{r_1}\cdots \xi^{r_p}
\eta^{i}\ ,\\[10pt]
\F^{(\ell,1/2)}&=&C^{(\ell,1/2)}+\pi\Big(C^{(\ell,1/2)}{}^\dagger\Big)
\star \C\ ,\\[10pt]
C^{(\ell,1/2)}&=&\sum_{n-m-p=4\ell+3}
\F^{(\ell,1/2)}_{r_1\dots r_p\,i}(m,n)\xi^{r_1}\cdots \xi^{r_p}
\eta^{i}\ ,\ee

where the scalar
field at level $(-1,1/2)$ obey the following reality condition

\be
\left(\F^{(-1,1/2)}_{r,i}(m,m)\right)^\dagger=(-1)^{m+1}\e_{rs}\e_{ij}
\F^{(-1,1/2)}_{s,j}(m,m)\ .\la{hypreality}\ee

The fields at level $(-1,1/2)$ form a hypermultiplet with the
following expansion

\be C^{(-1,1/2)}=\x^r\eta^i\f_{r\,i}+\yb^{\ad}\eta^i\l_{\ad\,i}+\cdots\ ,
\la{hyper}\ee

where \eq{hypreality} implies the reality conditions $\f_{r\,i}=
\e_{rs}\e_{ij}(\f_{s\,j})^\dagger$.

In summary, in addition to the fields of the minimal theory, the
non-minimal theory contains an extra set of $U(1)_f$ charged
fields labelled by $(\ell,1/2)$ which at the lowest level contains
a hypermultiplet. Thus we can view the non-minimal theory as
coupling of the minimal theory to a representation of the minimal
algebra carried by the $j=1/2$ fields. Just as the $U(1)_R$
coupling is given by the ratio between Planck length and AdS
radius, so is the $U(1)_f$ coupling.

{\footnotesize
\tabcolsep=1mm
\begin{table}[t]
\bec
\begin{tabular}{|l|cccccccccl|}\hline
& & & & & & & & & & \\
{${(\ell,j)}\backslash s$} & $0$ & \ns{$\ft12$} & $1$
& \ns{$\ft32$} & $2$ & \ns{$\ft52$} &
$3$ & \ns{$\ft72$} & $4$ & $\cdots$
\\
& & & & & & & & & &  \\
\hline
& & & & & & & & & &  \\
$(-1,1/2)$ & $2\!+\!{\bar 2}$ & 2 & & & & & &
& &  \\
$(-1,1)$ & $1\!+\!{\bar 1}$ & $2$ & $1$ & & & &
& & &  \\
$(0,0)$ & & & $1$ & $2$ & $1$ & & & & &\\
$(0,1/2)$ & & & & $2$ & $4$ & $2$ & & &
& \\
$(0,1)$ & & & & & $1$ & $2$ & $1$ & & &\\
$(1,0)$ & & & & & & & $1$ & $2$ & $1$ &\\
\phantom{aa}$\vdots$ & & & & & & & & & &
 \\ \hline
\end{tabular}
\ec \caption{{\small The spectrum of massless physical fields of
the non-minimal $\cN=2$ theory based on $hs(2|4;1)$ arranged into
levels of $\cN=2$ supermultiplets labelled by $\ell$ and with
$s_{\rm max}=2\ell+2+j$ and $U(1)_f$ charge $2j(2-2j)$.}}
\label{spec2b}
\end{table}}


\section{$\cN=4$ Theories}


In this section we begin with the description of the minimal
$\cN=4$ theory \cite{kv1,vr1,vr2}, for which we also provide an
alternative formulation in the end of Section 5.2. We then extend
to non-minimal $\cN=4$ theories with extra $U(1)_f$ or $SU(2)_f$
generators. Their linearized field equations are given by
\eq{lcn12} and their full field equations are discussed in
Sections \ref{Sec:full} and truncations in \ref{Sec:cons}. In all
cases the physical field contents of the master fields are in
one-to-one correspondence with the spectra given in Tables
\ref{spec4}, \ref{spec4bb} and \ref{spec4c}, respectively.


\subsection{The Minimal $\cN=4$ Theory}


The minimal $\cN=4$ theory \cite{kv1} is based on a higher spin
algebra which we shall denote by\footnote{In \cite{kv1} the
algebra is denoted by $shs^E(4,4|0)\simeq husp(2;2|4)$.} $hs(4|4)$
whose maximal finite-dimensional subalgebra is given by

\be OSp(4|4)\subset hs(4|4)\ .\ee

The R-symmetry algebra is $SO(4)_R=SU(2)_+\times SU(2)_-$.
The $OSp(4|4)$ singleton is given by

\be \la{xi4} \Xi=D(\ft12,0;2_+) + D(1,\ft12;2_-)\ .\ee

The spectrum of the $hs(4|4)$ theory, which is tabulated in Table \ref{spec4},
is given by the anti-symmetric
product of two $OSp(4|4)$ singletons

\be \la{eqspec4} {\rm
Spectrum}\left[hs(4|4)\right]=\left(\Xi\otimes\Xi \right)_{\rm
A}.\ee

The
supergravity multiplet is now contained in the anti-symmetric product,
unlike all other cases discussed in this paper where it arises in the
symmetric product.

The minimal $\cN=4$ theory is obtained by starting from the
bosonic theory described in Section 2 and adding the real
fermionic oscillators $\xi^r$ $(r=1,\dots,4)$ obeying

\be \xi^r\star \xi^s=\xi^r\xi^s+\d^{rs}\ ,\ee

where $r$ labels an $SO(4)_R$ vector.
The master fields are defined as in \eq{an1}, \eq{fn1} and \eq{grassmann}
where
$\t$, $\pi$ and $\bar \pi$ are defined as in \eq{tau1}, \eq{pi1} and
\eq{pibar1} and

\be \C=\x^1\x^2\x^3\x^4\ .\ee

The master fields can be expanded as

\be A_\m=\sum_{\ell=0}^\infty A_\m^{(\ell)}\ ,
\qquad \F=\sum_{\ell=0}^\infty \F^{(\ell)}\ ,
\ee

where $A_\m^{(\ell)}$ and $\F^{(\ell)}$ are given by ($\ell\geq 0$)

\be \la{aell4} A_\m^{(\ell)}&=&\sum_{m+n+p=4\ell+2}
A^{(\ell)}_{\m,r_1\dots r_p}(m,n)\xi^{r_1}\cdots \xi^{r_p}\ ,\\[10pt]
\la{fell4} \F^{(\ell)}&=&C^{(\ell)}+\pi
\Big(C^{(\ell)}{}^\dagger\Big)\star \C\ ,\\[10pt]
C^{(\ell)}&=&\sum_{n-m-p=4\ell}
\F^{(\ell)}_{r_1\dots r_p}(m,n)\xi^{r_1}\cdots \xi^{r_p}\ .\ee

The gauge field $A^{(0)}_\m$ gauges the $OSp(4|4)$ generators
$P^{\a\ad}$, $M^{\a\b}$, $Q^r_\a= y_\a\x^r$ and

\be T^{rs}_\pm=\ft12 (1\pm \C)\star\x^r\x^s\ .\ee

The $SU(2)_\pm$ gauge couplings
have equal strength given by the higher spin gauge coupling. In this
sense the minimal $\cN=4$ theory is a higher spin extension of the
$SO(4)$ gauged version \cite{dfr} of the globally $SO(4)$ invariant $\cN=4$
supergravity. Although this model allows an extension
with independent $SU(2)_\pm$ gauge couplings \cite{gz}, it is not
clear whether this model can be extended to higher spin gauge theory.
There also exist an $SU(2)\times SU(2)$ gauged version
(with independent gauge couplings) \cite{fs} of the globally
$SU(4)$ invariant $\cN=4$ supergravity, which however does
not admit any AdS$_4$ vacuum, and therefore presumably cannot
be extended to a $\cN=4$, $D=4$ higher spin gauge theory.

The unitary Fock space realization of $hs(4|4)$ is obtained from
\eq{bososc} and the fermionic creation operators

\be \psi^{1\,\dagger}=\xi^1+i\xi^2\ ,\qquad \psi^{2\,\dagger}
=\xi^3+i\xi^4 .\ee

The singleton $\Xi$ is isomorphic to the space of states built
from an even number of oscillators. Since $hs(4|4)$ consists of
even polynomials in oscillators, it follows that $\Xi$ and hence
$(\Xi\otimes \Xi)_{\rm A}$ are UIRs of $hs(4|4)$.

{\footnotesize
\tabcolsep=1mm
\begin{table}[t]
\bec
\begin{tabular}{|c|clllllllllll|}\hline
& & & & & & & & & & & & \\
{$\ell\backslash s$} & $0$ & \ns{$\ft12$} & $1$
& \ns{$\ft32$} & $2$ & \ns{$\ft52$} &
$3$ & \ns{$\ft72$} & $4$ & \ns{$\ft92$} & $5$ & $\cdots$
\\
& & & & & & & & & & & & \\
\hline
& & & & & & & & & & & & \\
$0$ & $1\!+\!{\bar 1}$ & $4$ & $6$ & $4$ & $1$ & &
& & & & & \\
$1$ & & & & & $1$ & $4$ & $6$ & $4$ & $1$
& & & \\
$\vdots$ & & & & & & & & & & & &
 \\ \hline
\end{tabular}
\ec \caption{{\small The spectrum of massless physical fields of
the minimal $\cN=4$ theory based on $hs(4|4)$ arranged into levels
of $\cN=4$ supermultiplets labelled by $\ell$ and with $s_{\rm
max}=2\ell+2$.}} \label{spec4}
\end{table}}


\subsection{\la{Sec:alt} Non-minimal $\cN=4$ Theory with $SU(2)_f$
Flavor}


The minimal $\cN=4$ theory described in Section 5.1 can be
embedded as the neutral sector of a non-minimal $\cN=4$ theory
which is based on the higher spin algebra $hs(4|4;2)$ with maximal
finite-dimensional subalgebra

\be OSp(4|4)\times SU(2)_f\subset hs(4|4;2)\ .\ee

The extended spectrum is given by

\be \la{spec4b} {\rm Spectrum}[hs(4|4;2)]=(\Xi'\otimes\Xi')_{\rm
S}\ ,\ee

which is listed in Table \ref{spec4bb}, where the $SU(2)_f$ flavored
singleton is given by

\be \la{xip} \Xi'= D(\ft12,0;2,1,2)+D(1,\ft12;1,2,2)\ ,\ee

where the last arguments denote $SU(2)_f$ doublets. The oscillator
realization of $hs(4|4;2)$ is obtained by starting from the
oscillator realization of $hs(4|4)$ and adding four extra real
fermionic oscillators $\eta^i$ obeying \eq{fosc2} for $r,i=1,\dots,4$.
The $hs(4|4;2)$ master fields are defined by \eq{an1}, \eq{fn1}
and \eq{grassmann}
where now

\be \C=\x^1\x^2\x^3\x^4\ ,\ee

and by imposing the following projection

\be \C_\eta\star A=A\star \C_\eta=A\ ,\qquad
\C_\eta\star \F=\F\star \C_\eta=\F\ ,\la{gammaetaproj}\ee

where $\C_\eta=\eta^1\eta^2\eta^3\eta^4$.
The resulting expansions of the master fields are given by

\be A_\m&=&A_\m^{(-1,1)}+\sum_{\ell=0}^\infty \left(A_\m^{(\ell,0)}+
A_\m^{(\ell,1)}\right)\ ,\\
\F&=&\F^{(-1,1)}+\sum_{\ell=0}^\infty \left(\F^{(\ell,0)}+
\F^{(\ell,1)}\right)\ ,\ee

where

\be A_\m^{(\ell,0)}=A_\m^{(\ell)}\star(1+\C_\eta)\ ,\qquad \F^{(\ell,0)}=
\F^{(\ell)}\star(1+\C_\eta)\ ,\ee

where $A_\m^{(\ell)}$ and $\F^{(\ell)}$ are
given in \eq{aell4} and \eq{fell4}, respectively, and

\be A_\m^{(\ell,1)}&=&\sum_{m+n+p=4\ell+4}
A^{(\ell,1)}_{\m\,r_1\dots r_p\,i_1i_2}(m,n)\xi^{r_1}\cdots \xi^{r_p}
\eta^{i_1}\eta^{i_2}\star(1+\C_\eta)\ ,\\[10pt]
\F^{(\ell,1)}&=&C^{(\ell,1)}+\pi
\Big(C^{(\ell,1)}{}^\dagger\Big)\star \C
\ ,\\[10pt]
C^{(\ell,1)}&=&\sum_{n-m-p=4\ell+2}
\F^{(\ell,1)}_{r_1\dots r_p\,i_1i_2}(m,n)\xi^{r_1}\cdots \xi^{r_p}
\eta^{i_1}\eta^{i_2}\star(1+\C_\eta)\ ,\ee

where the scalars at level
$(-1,1)$ obey the following reality condition:

\be \left(\F^{(-1,1)}_{rs,ij}(m,m)\right)^\dagger=
\ft12 {(-1)^{m}} \e_{rstu}\F^{(-1,1)}_{tu,ij}(m,m)\ .\la{vectorreal}\ee

The
$SU(2)_f$ Yang-Mills multiplet at level $(-1,1)$ arises in the
expansion of $\F$ as

\be C^{(-1,1)}=\left(\x^r\x^s\f_{rs\,ij}+\yb^{\ad}\x^r\l_{\ad\,r\,ij}
+\yb^{\ad}\yb^{\bd}F_{\ad\bd\,\,ij}+\cdots\right)K^{ij}\ ,\la{vectorexp}
\ee

where the $SU(2)_f$ subalgebra generators are given by

\be K_{ij} = \eta_i \eta_j\star(1+\C_\eta)\ ,\la{su2f}\ee

and the scalars obey the reality condition $\f_{rs\,ij}=
\ft12 \e_{rstu}(\f_{tu\,ij})^\dagger$ which follows from \eq{vectorreal}.

The gauge field $A_\m^{(0,0)}$ gauges the $OSp(4|4)$ subalgebra and the
$A_\m^{(-1,1)}$ gauges the $SU(2)_f$ subalgebra.
The non-minimal $hs(4|4;2)$ theory can be viewed as
a coupling of the minimal theory to a representation of the minimal algebra
carried by the $j=1$ fields. We note that at level $(0,1)$ there
is a set of spin
$1$ gauge fields which gauge
the generators

\be L_{ij}=\C K_{ij}\ ,\la{su2ff}\ee

which schematically obey $[L,L]=K$ and $[K,L]=L$. Since $L_{ij}$ commuted
with the supersymmetry generators yield higher spin generators, the
$L_{ij}$ generators do not belong to the maximal finite-dimensional
subalgebra of $hs(4|4;2)$.

The unitary Fock space realization of $hs(4|4;2)$ is
obtained from \eq{bososc} and the fermionic creation operators ($a=1,\dots,4$)

\be \psi^{a\,\dagger}=\xi^{a}+i\eta^{a}\ .\ee

The singleton $\Xi'$ in \eq{xip} is the space of states that are
created by an even number of oscillators and that are annihilated
by $(1-\C_\eta)$.

Starting from the $hs(4|4;2)$ theory one can obtain an
alternative formulation of the minimal $hs(4|4)$ theory discussed in the
previous section. To do so one imposes

\be [K_{ij},A]_\star=0=[K_{ij},\F]_\star\ .\la{altn4}\ee

The minimal singleton $\Xi$ given in \eq{xi4}
and the minimal spectrum \eq{eqspec4} are embedded
into $\Xi'$ given in \eq{xip} and the non-minimal spectrum
\eq{spec4b}, respectively, as $SU(2)_f$
invariant subspaces:

\be \Xi&=&\Xi'/K_{ij}\ ,\qquad \\[10pt]
{\rm Spectrum} \left[hs(4,4)\right]&=&(\Xi'\otimes
\Xi')_S/K_{ij}\simeq (\Xi\otimes \Xi)_A\ .\ee

{\footnotesize
\tabcolsep=1mm
\begin{table}[t]
\bec
\begin{tabular}{|l|clllllllllll|}\hline
& & & & & & & & & & & & \\
{$(\ell,j)\backslash s$} & $0$ & \ns{$\ft12$} & $1$
& \ns{$\ft32$} & $2$ & \ns{$\ft52$} &
$3$ & \ns{$\ft72$} & $4$ & \ns{$\ft92$} & $5$ & $\cdots$
\\
& & & & & & & & & & & & \\
\hline
& & & & & & & & & & & & \\
$(-1,1)$ & $6^3$ & $4^3$ & $1^3$ & & & & & &
& & & \\
$(0,0)$ & $1\!+\!{\bar 1}$ & $4$ & $6$ & $4$ & $1$ &
& & & & & & \\
$(0,1)$ & & & $1^3$ & $4^3$ & $6^3$ & $4^3$ & $1^3$
& & & & & \\
$(1,0)$ & & & & & $1$ & $4$ & $6$ & $4$ &
$1$ & & & \\
$(1,1)$ & & & & & & & $1^3$ & $4^3$ & $6^3$
& $4^3$ & $1^3$ &  \\
\phantom{aa}$\vdots$ & & & & & & & & & & & &
 \\ \hline
\end{tabular}
\ec \caption{{\small The spectrum of massless physical fields of
the non-minimal $\cN=4$ theory based on $hs(4|4;2)$ arranged into
levels of $\cN=4$ supermultiplets labelled by $(\ell,j)$ and with
$s_{\rm max}=2\ell+2+j$. The superscripts denote $SU(2)_f$
triplets. }} \label{spec4bb}
\end{table}}


\subsection{Non-minimal $\cN=4$ Theory with $U(1)_f$ Extension}


{\footnotesize
\tabcolsep=1mm
\begin{table}[h!]
\bec
\begin{tabular}{|l|clllllllllll|}\hline
& & & & & & & & & & & & \\
{$(\ell,j)\backslash s$} & $0$ & \ns{$\ft12$} & $1$
& \ns{$\ft32$} & $2$ & \ns{$\ft52$} &
$3$ & \ns{$\ft72$} & $4$ & \ns{$\ft92$} & $5$ & $\cdots$
\\
& & & & & & & & & & & & \\
\hline
& & & & & & & & & & & & \\
$(-1,1)$ & $6$ & $4$ & $1$ & & & & & &
& & & \\
$(0,0)$ & $1\!+\!{\bar 1}$ & $4$ & $6$ & $4$ & $1$ &
& & & & & & \\
$(0,1)$ & & & $1$ & $4$ & $6$ & $4$ & $1$
& & & & & \\
$(1,0)$ & & & & & $1$ & $4$ & $6$ & $4$ &
$1$ & & & \\
$(1,1)$ & & & & & & & $1$ & $4$ & $6$
& $4$ & $1$ &  \\
\phantom{aa}$\vdots$ & & & & & & & & & & & &
 \\ \hline
\end{tabular}
\ec \caption{{\small The spectrum of massless physical fields of
the non-minimal $\cN=4$ theory based on $hs_0(4|4;1)$ arranged
into levels of $\cN=4$ supermultiplets labelled by $(\ell,j)$ and
with $s_{\rm max}=2\ell+2+j$.}} \label{spec4c}
\end{table}}

Starting from the $hs(4|4;2)$ theory described above
and imposing invariance under the
$U(1)_f$ generator $K_{12}$ defined in \eq{su2f} as

\be [K_{12},A]_\star=0=[K_{12},\F]_\star\ ,\la{k12cond}\ee

we obtain
a non-minimal $\cN=4$ theory based on a higher spin algebra $hs_0(4|4;1)$
with maximal finite-dimensional subalgebra

\be OSp(4|4)\times U(1)_f\subset hs_0(4|4;1)\ .\ee

The resulting spectrum is given by

\be {\rm Spectrum}\left[hs_0(4|4;1)\right]=(\Xi'\otimes \Xi')_S/K_{12}\simeq
\Xi\otimes\Xi\ ,\ee

where the singletons $\Xi'$ and $\Xi$ are given by \eq{xip} and
\eq{xi4}, respectively. The spectrum, which is given in Table \ref{spec4c},
decomposes into the $hs(4|4)$ spectrum $(\Xi\otimes \Xi)_{\rm A}$
given in Table \ref{spec4}, and $(\Xi\otimes \Xi)_{\rm S}$ which yields
the $j=1$ multiplets of Table \ref{spec4c}.
In particular, the lowest $j=1$ multiplet is an $\cN=4$
vector multiplet which is embedded into the $\F$ field
as in \eq{vectorexp} with $ij\ra 12$.

{\footnotesize \tabcolsep=1mm
\begin{table}[h!]
\bec
\begin{tabular}{|c|c|l|c|ll|l|}\hline
& & & & & & \\
Higher spin & Maximal finite-dim. & \phantom{aa}Fermionic
\phantom{aa}& \phantom{aa}Projection \phantom{aa}&&
\phantom{aaaaaa}Spectrum\phantom{aaaa}
& \phantom{aa}Singleton\phantom{aa} \\
algebra & subalgebra & \phantom{aa}oscillators & operator
&& &
\\
& & & && & \\ \hline & & && & &
\\
$hs(4)$ & $Sp(4)$ & \phantom{aa}-- & --
&&
$(D(\ft12,0)\otimes
D(\ft12,0)_{\rm S}$ & \phantom{aa}$D(\ft12,0)$
\\
& & & && & \\
$hs_0(4;1)$ & $Sp(4)\times U(1)_f$ & \phantom{aa}-- & --
&&
$D(\ft12,0)\otimes
D(\ft12,0)$ & \phantom{aa}$D(\ft12,0)$
\\& & && & &\\
\hline & & && & &
\\
$hs(1|4)$ & $OSp(1|4)$ & \phantom{aa}$\psi=\x+i\eta$ & --
&&
$(\Xi\otimes
\Xi)_{\rm S}$ & \phantom{aa}Eq. (\ref{singn1}) \\
& & & && & \\ \hline & & && & &
\\
$hs_0(2|4;1)$ & $OSp(2|4)\times U(1)_f$ &
\phantom{aa}$\psi^a=\x^{a}+i\eta^{a}$ & $K=\C_\eta$ && $(\Xi\otimes
\Xi)_{\rm S}/K$ & \phantom{aa}Eq. (\ref{xi2}) \\
& & \phantom{aa}$(a=1,2)$& && $\simeq (\Xi_1\otimes
\overline{\Xi}_{-1})_{\rm S}$ & \\
& & && & & \\
$hs(2|4;1)$ & $OSp(2|4)\times U(1)_f$ &
\phantom{aa}$\psi^a=\x^{a}+i\eta^{a}$ & -- && $(\Xi\otimes
\Xi)_{\rm S}$ & \phantom{aa}Eq. (\ref{xi2}) \\
& &\phantom{aa}$(a=1,2)$ && & & \\ & &
&& & & \\\hline & & && & & \\
$hs(4|4)$ & $OSp(4|4)$ & \phantom{aa}$\psi^a=\x^{a}+i\x^{a+2}$ &
-- && $(\Xi\otimes
\Xi)_{\rm A}$ & \phantom{aa}Eq. (\ref{xi4}) \\
& & \phantom{aa}$(a=1,2)$& && & \\
& & && & & \\
$hs(4|4;2)$ & $OSp(4|4)\times SU(2)_f$ &
\phantom{aa}$\psi^a=\x^{a}+i\eta^{a}$ & -- && $(\Xi'\otimes
\Xi')_{\rm S}$ & \phantom{aa}Eq. (\ref{xip}) \\
& &\phantom{aa}$(a=1,\dots,4)$& && & \\
& & && & & \\
$hs_0(4|4;1)$ & $OSp(4|4)\times U(1)_f$ &
\phantom{aa}$\psi^a=\x^{a}+i\eta^{a}$ &
$K_{12}=\eta_1\eta_2$ && $(\Xi'\otimes
\Xi')_{\rm S}/K_{12}$ & \phantom{aa}Eq. (\ref{xip}) \\
& &\phantom{aa}$(a=1,\dots,4)$ & &&$\simeq \Xi\otimes \Xi$
& \phantom{aa}Eq. (\ref{xi4})\\
& & & && & \\
$hs(4|4)$ & $OSp(4|4)$ & \phantom{aa}$\psi^a=\x^{a}+i\eta^{a}$ &
$K_{rs}=\eta_r\eta_s$ && $(\Xi'\otimes
\Xi')_{\rm S}/K_{rs}$ & \phantom{aa}Eq. (\ref{xip}) \\
& &\phantom{aa}$(a=1,\dots,4)$ & &&$\simeq (\Xi\otimes
\Xi)_{\rm A}$ &\phantom{aa}Eq. (\ref{xi4})\\
& & & && & \\ \hline& & & && & \\
$hs(8|4)$ & $OSp(8|4)$ & \phantom{aa}$\psi^a=\x^a+i\x^{a+4}$ & -- &&
$(\Xi\otimes\Xi)_{\rm S}$ & \phantom{aa}See \\
& &\phantom{aa}$(a=1,\dots,4)$ & && &\phantom{aa}Caption \\& & & && & \\ \hline
\end{tabular}
\ec \caption{{\footnotesize Summary of $\cN=0,1,2,4,8$ higher spin
algebras. $hs(4)$ and $hs_0(4;1)$ are bosonic. For the rest, the
first two arguments refer to $OSp(\cN|4)$ and the last argument
separated by a semicolon refers to additional flavor groups in the
maximal finite-dimensional subalgebra and the zero subscripts mean
that the spectrum is $U(1)_f$ neutral. The third column contains
the fermionic oscillators used in the master field expansions and
the unitary oscillator realizations. The fifth column contains the
singleton product annihilated by the operators listed in the
fourth column. The corresponding conditions on the master fields
are summarized in \eq{etacond}. The $OSp(8|4)$ singleton is given
by $\Xi=D(\ft12,0;8_s)+D(1,\ft12;8_c)$. The minimal
${\cN}=0,1,2,4,8$ algebras have been given previously in the
literature \cite{kv0,kv1}, using a different notation as follows:
$hs(4)\simeq hs_2(1)$, $hs(1|4)\simeq shs^f(1|0)$,
$hs_0(2|4;1)\simeq shs^E(2,4)\simeq hu(1;1|4)$, $hs(4|4)\simeq
shs^E(4,4|0)\simeq husp(2;2|4)$ and $hs(8|4)\simeq shs^E(8,4|0)
\simeq ho(8;8|4)$. }} \label{summary}
\end{table}}


\section{\la{Sec:full}The Full Field Equations}


The general method for formulating full higher spin field
equations in $D=4$ has been developed by Vasiliev
\cite{4dv0,4dv1}. In particular, this formalism yields the field
equations for the minimal $\cN=0$ mod $4$ theories based on the
minimal $\cN=0,4,8$ algebras \cite{kv0,kv1} which we have reviewed
here; see Table \ref{summary}. Utilizing the basic structure of
the $\cN=0$ mod $4$ theories, which has been studied in detail for
$\cN=8$ in \cite{us3,us4} and for $\cN=0$ in \cite{analysis}, we
here express the full field equations for all the specific models
considered in this paper in a universal form. This is possible
due to the universal form of the
`twisted' reality condition \eq{eq:conboson2}
on the scalar master field $\Phi$. It is worth
pointing out that this universal form does not require Kleinian
operators \cite{mat,kv0,kv1,valg,vr2,vpro}. Hence the extra
auxiliary fields that arise for certain $\cN$ upon introducing
Kleinian operators, along the lines explained in
\cite{mat,vpro,vr2}, do not arise in the formulation presented
here. In particular, this yields minimal $\cN=1,2$ theories based
on $hs(1|4)$ and $hs_0(2|4;1)$, which to our best knowledge have
not been given previously. It remains an open problem, however, to
determine whether these $\cN=1,2$ theories can be obtained from
the `Kleinian' $\cN=1,2$ theories (with extra auxiliary fields) by
means of consistent truncations. This point will be discussed
further in the conclusion of this section.

Thus, following \cite{4dv1} we extend the ordinary spacetime by
non-commutative internal coordinates $z^{\a}$ and
$\zd=(z^\a)^\dagger$. We then consider Grassmann even master
fields $\widehat A=dx^\m \Ah_\m+dz^\a\Ah_\a+ d\zb^{\ad}\Ah_{\ad}$
and $\widehat \F$ obeying

\be
\la{tha}\tau(\Ah)&=&-\Ah\ ,\qquad~~ \Ah^{\dagger}~=~-\Ah\ ,\\[10pt]
\la{thf} \tau(\widehat \F)&=&\pb(\widehat \F)\ ,\qquad
\widehat \F^{\dagger}~=~\pi(\widehat \F)\star \C\ ,\ee

where $\t$, $\pi$ and $\bar \pi$ are defined by

\be
\tau(f(y,\yb,\x,\eta,z,\zb))&=&f(iy,i\yb,i\x,-i\eta,-iz,-i\zb)\ ,\\
\pi(f(y,\yb,\x,\eta,z,\zb))&=&f(-y,\yb,\x,\eta,-z,\zb)\ ,\\
\pb(f(y,\yb,\x,\eta,z,\zb))&=&f(y,-\yb,\x,\eta,z,-\zb)\ , \ee

and that their action commutes with the exterior derivative $d$ on
the $(x,z,\bar z)$ space. In \eq{tha} and \eq{thf} the appropriate
$\x$ and $\eta$ oscillators and $\C$ operators have been defined
in the previous sections ($\C=1$ in the minimal bosonic theory).
The master fields also obey the various model dependent algebraic
reduction conditions on the $\eta$ dependence which have been
discussed case by case in the previous sections, and which can be
summarized as follows:

\be \ba{lclcllcl} hs_0(2|4;1)&:& [\C_\eta,\Ah]_\star&=&0\ ,&
[\C_\eta,\Fh]_\star&=&0\ ,\\
[10pt]
hs_0(4|4;1)&:& \C_\eta\star \Ah&=&\Ah\star \C_\eta=\Ah\ ,\phantom{aaa}&
\C_\eta\star\Fh&=&\Fh\star\C_\eta=\Fh\ ,\\
&& [K_{12},\Ah]_\star&=&0\ ,& [K_{12},\Fh]_\star&=&0\ ,\\[10pt]
hs(4|4;2)&:& \C_\eta\star \Ah&=&\Ah\star \C_\eta=\Ah\ ,&
\C_\eta\star\Fh&=&\Fh\star\C_\eta=\Fh\ .\ea\la{etacond}\ee

The conditions \eq{tha} and \eq{thf} and the conditions \eq{Gamma} on the
$\C$-operator are the same as those defining
the minimal $\cN=8$ model \cite{4dv1,kv1}. Hence the structure of the
full $\cN=8$ field equations, which
were studied in detail in \cite{us4},
carry over to all the specific models considered in this paper.
Moreover, we observe that the reduction conditions \eq{etacond},
which do not arise in the
$\cN=8$ theory, are naturally consistent with the full field equations.

Thus, the full field equations
for the models considered in this paper follow from the constraints

\be \la{bm1}{\widehat F}&=&\ft{i}4dz^{\a}\wedge
dz_{\a}\cV(\Fh\star\kappa\C)+ \ft{i}4d\zb^{\ad}\wedge
d\zb_{\ad}\overline{\cV}(\Fh\star\kappa^\dagger)\
,\\[10pt]
\la{bm2}{\widehat D}\Fh&=&0\ ,\ee

subject to the initial conditions

\be \la{eq:initial}
\Ah_{\m}{\Big |}_{Z=0}=A_{\m}\ ,\qquad \Ah_{\a}{\Big |}_{Z=0}=0\ ,\qquad
\Fh{\Big |}_{Z=0}=\F\ ,\ee

where

\be \widehat F&=&
d \Ah+\Ah\wedge\star \Ah\ ,\\[10pt]
{\widehat D}\Fh&=&d \Fh+\Ah\star\Fh-\Fh\star\pb(\Ah)\ ,\\[10pt]
\kappa&=&\exp iz_{\a}y^{\a}\ .
\ee

The quantity $\cV(X)$ in \eq{bm1} is an odd $\star$-function

\be \cV(-X)=-\cV(X)\ ,\la{vodd}\ee

where $\cV(X) = b_1 X + b_3 X\star X\star X+\cdots$ for some
complex constants $b_n$, and

\be (\cV(X))^\dagger=\overline{\cV}(X^\dagger)\ .\ee

As was first shown by Vasiliev \cite{vr2}, local Lorentz
invariance severely constrains the form of the spinorial
components of $\widehat F$. Local Lorentz invariance combined with
\eq{tha} and \eq{thf} implies $\widehat F_{\a\ad}=0$ \cite{vr2}
and $\cV(X)=-\cV(-X)$ \cite{analysis}.

The equations in \eq{bm1} and \eq{bm2} are integrable and invariant
under

\be \d {\widehat A}&=&d {\widehat \e}+\Ah\star {\widehat \e}-{\widehat
\e}\star\Ah\
,\\
\d\Fh&=&\Fh\star\pb({\widehat \e})-{\widehat \e}\star\Fh\ , \ee

where ${\widehat \e}$ obeys \eq{tha}. As shown in \cite{4dv1},
one can use the $Z$-components of \eq{bm1} and \eq{bm2} to solve for the $Z$
dependence by expanding in powers of $\F$, and obtain
the higher spin field equations upon restricting the remaining
spacetime components of \eq{bm1} and \eq{bm2} to $Z_{\underline
\a}=0$. In particular, this analysis yields the
linearized field equations \eq{lcn12}.
We also note that the detailed analysis of the $\F$-expansion
given in \cite{analysis}
carry over straightforwardly to the cases of $\cN=1,2,4$.

The presence of the function $\cV(X)$ means that the
gauge invariant higher spin interactions are not unique. This
ambiguity underlines the fact that the
massless higher spin gauge theory should be thought of as
(a consistent truncation of) some effective description
of some underlying fundamental theory, perhaps along the lines
discussed in \cite{holo}, which of course requires a
particular form of $\cV(X)$ that should be
predicted by the underlying theory.

Importantly, the general form of the interaction ambiguity is more
constrained in the present formulation than in the formulation
involving non-trivial Kleinian operators and extra auxiliary
fields. In the latter case the interaction ambiguity is described
by an arbitrary function \cite{4dv1,vr2,vpro}, with both even and
odd powers of the scalar master field. As indicated in \cite{vr2},
it is possible, however, to consistently truncate the auxiliary
fields in a certain class of models, provided that the interaction
ambiguity is described by an odd function. Whether this leads
to the universal formulation considered here remains an open problem.


\section{\la{Sec:cons}Consistent Truncations}


\begin{figure}
\begin{picture}(100,170)(-110,0)

\put(-70,160){\makebox(0,0){$\cN=8$:}}
\put(-70,120){\makebox(0,0){$\cN=4$:}}
\put(-70,80){\makebox(0,0){$\cN=2$:}}
\put(-70,40){\makebox(0,0){$\cN=1$:}}
\put(-70,0){\makebox(0,0){$\cN=0$:}}

\put(0,140){\makebox(0,0){\vector(0,-1){20}}}
\put(100,100){\makebox(0,0){\vector(0,-1){20}}}
\put(200,100){\makebox(0,0){\vector(0,-1){20}}}
\put(200,40){\makebox(0,0){\vector(0,-1){60}}}
\put(300,20){\makebox(0,0){\vector(0,-1){20}}}

\put(50,125){\makebox(0,0){\vector(1,0){30}}}
\put(150,125){\makebox(0,0){\vector(1,0){30}}}
\put(150,85){\makebox(0,0){\vector(1,0){30}}}
\put(250,5){\makebox(0,0){\vector(1,0){30}}}

\put(248,56){\makebox(0,0){\vector(2,-1){60}}}

\put(0,160){\makebox(0,0){$hs(8|4)$}}

\put(0,120){\makebox(0,0){$hs(4|4;2)$}}
\put(100,120){\makebox(0,0){$hs_0(4|4;1)$}}
\put(200,120){\makebox(0,0){$hs(4|4)$}}

\put(100,80){\makebox(0,0){$hs(2|4;1)$}}
\put(200,80){\makebox(0,0){$hs_0(2|4;1)$}}

\put(300,40){\makebox(0,0){$hs(1|4)$}}

\put(200,0){\makebox(0,0){$hs_0(4;1)$}}
\put(300,0){\makebox(0,0){$hs(4)$}}

\end{picture}
\caption{{\small The arrows denote consistent truncations of
higher spin gauge theories described in detail in Section 7. }}
\la{fig1}
\end{figure}

Starting from the minimal $\cN=8$ model based on the higher
spin algebra $hs(8|4)$, it is possible to obtain
the models described in this paper by means of the consistent truncations
summarized in Figure \ref{fig1}. Throughout this section {\it all} master
fields are understood to be hatted, and we have dropped the hats for
simplicity.\\

$\bullet$ $\  hs(8|4)\ra hs(4|4;2)$:

The consistent truncation $hs(8|4)\ra hs(4|4;2)$ is obtained by
setting equal to zero all components of the $hs(8|4)$ master fields
$A$ and $\F$ with an odd number of $\x^{5,6,7,8}$ oscillators followed by
a multiplication with the projector $\ft12(1+\x^5\x^6\x^7\x^8)$.
Furthermore, identifying $\x^{5,6,7,8}$ with the
$\eta^{1,2,3,4}$ oscillators in
the $hs(4|4;2)$ model, the reduced $hs(8|4)$ master
fields become $hs(4|4;2)$ master fields. Inserting the reduced
$hs(8|4)$ master fields into the full $hs(8|4)$ constraints
\eq{bm1} and \eq{bm2} and noting that

\be (1+\x^5\x^6\x^7\x^8)\C|_{hs(8|4)}\ra (1+\C_\eta)\C|_{hs(4|4;2)}\ ,\ee

we obtain consistently full $hs(4|4;2)$ master constraints \eq{bm1}
and \eq{bm2} with
the same $\cV$.\\

$\bullet$ $\  hs(4|4;2)\ra hs_0(4|4;1)\ra hs(4|4)$:

The truncation $hs(4|4;2)\ra hs_0(4|4;1)$ amounts to imposing
neutrality under $U(1)_f\subset SU(2)_f$, as described in Section
5.3~. As a result, the $SU(2)_f$ triplets residing at levels $(\ell,1)$
in Table \ref{spec4bb}
are replaced by $U(1)_f$ singlets as shown in Table \ref{spec4c}. The
truncation $hs_0(4|4;1)\ra hs(4|4)$ is achieved by dropping
the terms in the $hs_0(4|4;1)$ master fields which contain
the $U(1)_f$ generator, i.e. the $j=1$ multiplets of Table \ref{spec4c}.
Both of these truncations are consistent with the master equations
\eq{bm1} and \eq{bm2}.\\

$\bullet$ $\  hs(4|4)\ra hs_0(2|4;1)$:

To describe the consistent truncation $hs(4|4)\ra hs_0(2|4;1)$, we
begin by writing the $hs(4|4)$ master fields as

\be A&=& A^{(0)} +\x^r \Psi^{(0)}_r+\x^r\x^s A^{(0)}_{rs}+
\x^r\star\C_\x\Psi^{(1)}_r
+\C_\x A^{(1)}\ ,\\
\F&=& \F^{(0)} +\x^r \chi^{(0)}_r+\x^r\x^s\F^{(0)}_{rs}+
\x^r\star\C_\x\chi^{(1)}_r +\C_\x\F^{(1)}\ .\ee

We then set

\be
\Psi^{(0)}_{r}&=&\Psi^{(0)}_{r}=0\ \mbox{ for }r=3,4\la{red1}\\
\chi^{(0)}_r&=&\chi^{(0)}_r=0\ \mbox{ for }r=3,4 \la{red11}\\[10pt]
A^{(0)}_{12}&\equiv& B^{(0)}\ ,\qquad A^{(0)}_{34}\equiv
B^{(1)}\ ,\la{red2} \\ \F^{(0)}_{12}&\equiv&
\varphi^{(0)}\ ,\qquad
A^{(0)}_{34}\equiv \varphi^{(1)}\ ,\la{red3} \\
A^{(0)}_{rs}&=&0\ ,\qquad\ \quad\F^{(0)}_{rs}=0\qquad\mbox{ for $rs\neq 12$ or
$34$}\ .\la{red4}\ee

Upon identifying the $\x^{3,4}$ oscillators in the
$hs(4|4)$ model with the $\eta^{1,2}$ oscillators
in the $hs_0(2|4;1)$ model, the reduced $hs(4|4)$ master
fields become the $hs_0(2|4;1)$ master fields:

\be \la{n2ma} A_{\rm red}&=& A^{(0)} +\x^r
\Psi^{(0)}_r+\C_\x B^{(0)}+\C_\eta
B^{(1)}+ \x^r\C_\eta\Psi^{(1)}_r +\C A^{(1)}\ ,\\
\la{n2mf} \F_{\rm red}&=& \F^{(0)} +\x^r
\chi^{(0)}_r+\C_\x\varphi^{(0)}+
\C_\eta\varphi^{(1)}+ \x^r\C_\eta\chi^{(1)}_r
+\C\F^{(1)}\ ,\ee

where now $r=1,2$, $\C_\x=i\x^1\x^2$,
$\C_\eta=i\eta^1\eta^2$, $\C=\C_\x\C_\eta$ and
we have absorbed $\e_{rs}$ into redefinitions of
$\Psi^{(1)}_r$ and $\chi^{(1)}_r$. Inserting
the reduced $hs(4|4)$ master fields into the full $hs(4|4)$
constraints \eq{bm1} and \eq{bm2} and noting that

\be \C|_{hs(4|4)}\equiv
\C_\x \ra \C_\x\C_\eta\equiv \C|_{hs_0(2|4;1)}\ ,\ee

we obtain consistently the
full $hs_0(2|4;1)$ constraints.\\

$\bullet$ $\ hs_0(4|4;1)\ra hs(2|4;1)$:

To achieve the consistent truncation $hs_0(4|4;1)\ra hs(2|4;1)$
we begin by writing we the $hs_0(4|4;1)$ master
fields as

\be \nn A&=&A^{(0)} +\x^r
\Psi^{(0)}_r+\x^r\x^sA^{(0)}_{rs}+
\x^r\star\C_\x\Psi^{(1)}_r
+\C_\x A^{(1)}\\
&&+
\eta_1\eta_2\left(A^{(2)}+\x^r\Psi^{(2)}_r+\x^r\x^s
A^{(2)}_{rs}+\x^r\star\C_\x\Psi^{(3)}_r+
\C_\x A^{(3)}\right)\
,\\[10pt] \nn
\F&=& \F^{(0)} +\x^r
\chi^{(0)}_r+\x^r\x^s\F^{(0)}_{rs}+
\x^r\star\C_\x\chi^{(1)}_r +\C_\x\F^{(1)}\\ &&
 +\eta_1\eta_2\left(\F^{(2)} +\x^r
\chi^{(2)}_r+\x^r\x^s\F^{(2)}_{rs}+
\x^r\star\C_\x\chi^{(3)}_r
+\C_\x\F^{(3)}\right)\ ,\ee

where we have taken into account the $\C_\eta$ projection in
\eq{gammaetaproj}.
We next impose \eq{red1},
\eq{red2}, \eq{red3}, \eq{red4} and set

\be
\Psi^{(2)}_{r}&=&
\Psi^{(3)}_{r}=0\
\mbox{ for }r=3,4 \\
\chi^{(2)}_r&=&\chi^{(3)}_r=0
\mbox{ for }r=3,4\ ,\\[10pt]
A^{(2)}&=&A^{(3)}=0\ ,\\
A^{(2)}_{12}&=&A^{(2)}_{34}=0\ ,\\
\F^{(2)}_{12}&=&\F^{(2)}_{34}=0\ .\ee

Upon identifying $\eta_1\eta_2\x^{3,4}$ in the
$hs_0(4|4;1)$ model with $\eta^{1,2}$
in the $hs(2|4;1)$ model, the reduced
master fields become the $hs(2|4;1)$ master fields. This leads to
a consistent truncation of \eq{bm1} and \eq{bm2}.\\

$\bullet$ $\ hs_0(2|4;1)\ra hs(1|4)\ra hs(4)$:

To describe the consistent truncation $hs_0(2|4;1)\ra hs(1|4)$,
we start from the $hs_0(2|4;1)$ master fields as written in \eq{n2ma} and
\eq{n2mf} and set

\be \Psi^{(0)}_1\equiv \Psi^{(0)}\ ,\quad
\Psi^{(1)}_2\equiv\Psi^{(1)}\ ,\quad
\chi^{(0)}_1&\equiv& \chi^{(0)}\ ,\quad
\chi^{(1)}_2\equiv
\chi^{(1)}\ ,\ee \be
\Psi^{(0)}_2=
\Psi^{(1)}_1=
\chi^{(0)}_2=
\chi^{(1)}_1 =0\ .\ee

Upon identifying $\x^1$ and $\C_\eta\x^2$ in the $hs_0(2|4;1)$ model
with the $hs(1|4)$ oscillators $\x$ and $\eta$, respectively, and
noting that

\be \C|_{hs_0(2|4;1)}\equiv \C_\x\C_\eta\ra i\x\eta\equiv
\C|_{hs(1|4)}\ ,\ee

the reduced $hs_0(2|4;1)$ master fields become the $hs(1|4)$ master fields

\be \la{n1ma} A_{\rm red}&=&A^{(0)}+\x\Psi^{(0)}+
\eta\Psi^{(1)}+\C A^{(1)}\
,\\
\la{n1mf} \F_{\rm red}&=&\F^{(0)}+\x\chi^{(0)}+
\eta\chi^{(1)}+\C\F^{(1)}\ ,\ee

where $\C=i\x\eta$ and this results in a consistent
truncation of \eq{bm1} and \eq{bm2}.
The further consistent truncation $hs(1|4)\ra hs(4)$ is obtained by
starting from \eq{n1ma} and \eq{n1mf} and setting
the fermions equal to zero and

\be A^{(0)}=A^{(1)}\ ,\qquad \F^{(0)}=\F^{(1)}\ .\ee

Inserting this into the $hs(1|4)$ master equation we obtain the
$hs(4)$ master equation multiplied by
$1+i\xi\eta$ that can be trivially dropped since the master fields no
longer depends on $\xi$ and $\eta$ oscillators.\\

$\bullet$ $\ hs_0(2|4;1)\ra hs_0(4;1)\ra hs(4)$:

The consistent truncation $hs_0(2|4;1)\ra hs_0(4;1)$
is obtained by starting from the $hs_0(2|4;1)$ master fields
as written in \eq{n2ma} and
\eq{n2mf} and setting the fermions equal to zero and

\be A^{(0)}=A^{(1)}&,&B^{(0)}=B^{(1)}\ ,\\
\F^{(0)}=\F^{(1)}&,&\varphi^{(0)}=\varphi^{(1)}\ .\ee

Dropping the overall $1+\C$ results in the following $hs_0(4;1)$ master fields

\be A&=& \left[A^{(0)}+\C_\x B^{(1)}\right]\star(1+\C)\ ,\la{alt1}\\
\F&=&\left[\F^{(0)}+\C_\x\varphi^{(1)}\right]\star(1+\C)\
,\la{alt2}\ee

obeying

\be \t(A^{(0)})&=&-A^{(0)}\ ,\qquad \t(B^{(1)})=B^{(1)}\
,\la{alt3}\\ \t(\F^{(0)})&=&\bar\pi(\F^{(0)}) \ ,\qquad
\t(\varphi^{(1)})=-\bar\pi(\varphi^{(1)})\ ,\la{alt4}\ee

and reality conditions as in \eq{tha} and \eq{thf}. The fields
with even spins
are contained in $A^{(0)}$ and $\F^{(0)}$ and those with odd spins
in $B^{(1)}$ and $\varphi^{(1)}$. Finally, the consistent truncation
$hs_0(4;1)\ra hs(4)$ is obtained by setting $B^{(1)}$ and $\varphi^{(1)}$
equal to zero.


\section{M Theory on $ AdS_4\times N^{010}$ and Matter Coupled
$\cN=4$ Higher Spin Gauge Theory}


The consistent truncation schemes described in the previous section
implies that the proposed holography between the $hs(8|4)$ higher
spin gauge theory and the $SU(N)_c$-invariant free $OSp(8|4)$
singleton SCFT in $d=3$
is testable using also models with $\cN<8$.
The $\cN<8$ theories based on the higher spin
algebras $hs(\cN|4)$ may also be relevant as truncations to
the massless sector of
unbroken phases of M theory on $AdS_4\times X^7$ for suitable
internal spaces $X^7$. All smooth $X^7$ with $\cN\geq 1$
that can arise in this way
have been classified \cite{warner}.
As in the case of the Type IIB theory on $AdS_5\times T^{11}$
\cite{klebanov},
the geometry of the conifold
${\cal C}^8(X^7)$ encodes a great deal of
information on the nature of the
holographic SCFT which describes the low energy
dynamics of M2 branes sitting at the apex of ${\cal C}^8$.

The possible $\cN\geq 1$ for smooth $X^7$ are $\cN=1,2,3,8$. While there exist
a number of ${\cN}=1,2$ compactifications, there is only one known
case with ${\cN}=3$, namely $AdS_4\times N^{010}$. The conic geometries
and associated SCFTs have been studied for $\cN=2$ \cite{n2,v52} and $\cN=3$
\cite{n3}. Due to its higher degree of symmetry and its uniqueness,
we shall focus our attention on the $\cN=3$ case in the rest of this section.

The manifold $N^{010}=SU(3)/U(1)$ has isometry group
$SU(3)\times SU(2)_R$. The spectrum of 11D supergravity on
$AdS_4\times N^{010}$ is known \cite{n010} and the only $4D$ massless fields in
the spectrum are the $\cN=3$ supergravity multiplet, a special vector
multiplet known as the Betti multiplet and an $SU(3)$ Yang-Mills multiplet.

The interacting SCFT has been proposed \cite{n3} to be the IR
fixed point of a $d=3$, $\cN=3$ gauged sigma model with field
content given by a vector multiplet in the adjoint representation
of a (color) gauge group $SU(N)_1\times SU(N)_2$ and an $SU(3)$
triplet of scalar multiplets transforming in the $SU(N)_1\times
SU(N)_2$ bi-fundamental representation.

In the UV limit there is an $\cN=3\ra 4$ enhancement and the
field theory is described by the following free $OSp(4|4)$ singleton
superfields \cite{n3}

\be Y^A{}_a{}^b\ ,\qquad Y^A{}_{a'}{}^{b'}\ ,\qquad
U^{A\,I}{}_a{}^{b'}\ ,\ee

where $A=1,2$ is an $SU(2)_+$ doublet index, $a,a'=1,\dots, N$ and
$I=1,\dots,3$ labels an $SU(3)$ triplet. The higher spin
symmetries of the free SCFT therefore form an algebra which is the
direct sum of an $hs(4|4;2)$ algebra made out of bilinears in $Y$
superfields, and an $U(3)$-flavored extension of $hs_0(4|4;1)$
\cite{kv1} made out of bilinears in the $U$ superfield. Thus the
full higher spin algebra is given by

\be hs[N^{010}] = hs(4|4;2) \oplus \left[hs_0(4|4;1)\otimes U(3)\right]
\ ,\ee

where an element $f^I{}_J\in hs_0(4|4;1)\otimes U(3)$ is defined by the
following conditions:

\be \t(f^I{}_J)=-f^I{}_J\ ,\qquad (f^I{}_J)^\dagger=-f^J{}_I\ .\ee

The $U(3)$-flavored higher spin algebra
is irreducible, since the commutator of two $SU(3)$-flavored
generators contains $U(1)\subset U(3)$ flavored generators. It is
consistent, however, to set the fields in the $SU(3)$-flavored
sector equal to zero.

Prior to breaking the higher spin symmetries down to those of
the supergravity theory on $AdS_4\times N^{010}$, the spectrum consists
of the $hs(4|4;2)$ spectrum given in Table \ref{spec4bb}
and a $U(3)$-valued version of the $hs_0(4|4;1)\otimes U(3)$ spectrum given in
Table \ref{spec4c}. Thus the resulting $s_{\rm max}=1,2$ multiplets in
the full spectrum are:

\be s_{\rm max}=1:&& (6\ 4\ 1\ )\qquad\quad\qquad\qquad\qquad
\mbox{from $hs_0(4|4;1)
\otimes U(3)$}\ ,\la{lm1}\\ && (6\ 4\ 1\ )\otimes 8 \ \quad\qquad\qquad\qquad
\mbox{from $hs_0(4|4;1)
\otimes U(3)$}\ ,
\la{lm2}\\[5pt]
&&(6^3\ 4^3\ 1^3\ )\qquad \ \ \,\quad\qquad\qquad\mbox{from $hs(4|4;2)$}\ ,
\la{lm3}\\[10pt]
s_{\rm max}=2:&& (1+\bar{1}\ 4\ 6\ 4\ 1)\qquad\qquad\qquad
\mbox{from $hs_0(4|4;1)
\otimes U(3)$}\ ,\la{lm4}\\ &&(1+\bar{1}\ 4\ 6\ 4\ 1)\otimes 8
\qquad\ \qquad\mbox{from $hs_0(4|4;1)
\otimes U(3)$}\ ,\la{lm5}\\[5pt] && (1+\bar{1}\ 4\ 6\ 4\ 1)
\qquad\qquad\qquad\mbox{from $hs(4|4;2)$}\ .\la{lm6}\ee

Switching on interactions in the boundary field theory breaks
$\cN=4\ra 3$ and leads to anomalies in the conservation laws for
the higher spin currents as well as some currents in the $s_{\rm
max}\le 2$ sector. The anomaly multiplets couple to Higgs fields
in the bulk which are eaten by the gauge fields to form massive
multiplets \cite{edseminar, holo}. Thus the $\cN=4$ higher spin
symmetries break down to those of the matter coupled $\cN=3$
supergravity.

We note that the $11D$ supergravity KK spectrum in the broken
phase does not have any massive $s_{\rm max}=2$ multiplet which is
$SU(2)_R$ singlet. Hence we expect that the $8$-plet of $s_{\rm
max}=2$ multiplets in \eq{lm5}, and a linear combination of the
$s_{\rm max}=2$ multiplets in \eq{lm4} and \eq{lm6} will be eaten
up by suitable higher spin multiplets. We also expect that the
other combination undergoes a super-Higgs effect in which it eats
a triplet of vector multiplets to give the $\cN=3$ supergravity
multiplet and a massive spin $s_{\rm max}=3/2$ multiplet, known as
the shadow of the supergravity multiplet \cite{shadow}. It is
tempting to identify the required Higgs vector multiplets with the
$SU(2)_f$ triplet of vector multiplets given in \eq{lm3}. The fate
of the latter multiplets and whether they can indeed be identified
with the Higgs vector multiplets requires further analysis of the
anomalies associated with symmetry breaking due to interactions in
the boundary gauged sigma model.

Finally, the vector multiplets in \eq{lm1} and \eq{lm2} are the
natural candidates for the Betti multiplet and the $SU(3)$
Yang-Mills multiplets which arise in the massless sector of the $11D$
supergravity KK spectrum. \\[10pt]

{\Large \bf Acknowledgements}

We thank the University of Groningen and one of us (E.S.) thanks
also the University of Uppsala, for kind hospitality and a
stimulating work atmosphere. We also thank P.S. Howe for useful
discussions. The work of E.S. is supported in part by NSF Grant
PHY-0070964.

\pagebreak

\end{document}